\documentclass[%
 reprint,
 amsmath,amssymb,
 aps,
 pra,
]{revtex4-2}

\usepackage{graphicx}
\usepackage{dcolumn}
\usepackage{bm}
\usepackage[hidelinks]{hyperref}
\usepackage{nicefrac}
\usepackage{float}

\usepackage[normalem]{ulem}
\usepackage{xcolor}

\begin{document}

\preprint{APS/123-QED}

\title{Feedback Cooling and Thermometry of a Single Trapped Ion Using a Knife Edge}

\author{Hans Dang$^{1,2,\bigstar}$}
\email{hans.dang@mpl.mpg.de}
\author{Sebastian Luff$^{1,2,\bigstar}$}
\email{sebastian.luff@mpl.mpg.de}
\author{Martin Fischer$^{2}$}
\author{Markus Sondermann$^{1,2,3}$}
\author{Gerd Leuchs$^{1,2,4}$}

\affiliation{$^1$Friedrich-Alexander-Universität Erlangen-Nürnberg (FAU), Department of Physics, Staudtstr. 7/B2, D-91058, Erlangen, Germany}
\affiliation{$^2$Max Planck Institute for the Science of Light, Staudtstr. 2, D-91058, Erlangen, Germany}
\affiliation{$^3$Friedrich-Alexander-Universität Erlangen-Nürnberg (FAU), Lehrstuhl für Experimentalphysik, Staudtstr. 7/B2, D-91058, Erlangen, Germany}
\affiliation{$^4$Department of Physics, University of Ottawa, Ottawa, {Ontario K1N 6N5}, Canada}
\affiliation{$^\bigstar$These authors contributed equally.}

\date{December 18, 2025}

\begin{abstract}
We report on a simple and easy to implement method of feedback cooling trapped ions to temperatures below those achievable using only Doppler cooling. Additionally, the feedback cooling results in significantly shorter cooling times. For selected parameters, we demonstrate cooling to temperatures below \(\hbar\Gamma/2 k_\mathrm{B}\). The motion of a single ion is monitored in real-time, allowing for the generation of a feedback signal that is applied to an auxiliary trap electrode. Motion detection is implemented by imaging the fluorescence photons emitted by the ion onto a knife edge and detecting the transmitted light, a method used so far to cool trapped nanoparticles. The intensity modulation of the fluorescence resulting from the ion motion is used to generate and apply the feedback signal and also to determine the ion temperature. While the method benefits from a high rate of detected scattered photons, which can be a challenge, and which we address by using a parabolic mirror for collecting the fluorescence, we expect the method to also be applicable when using lenses with moderate numerical apertures.
\end{abstract}

\maketitle

\section{{Introduction}}
The long coherence times and stability of trapped ions make them a suitable choice for the experimental realization of quantum computing, quantum information processing applications~\cite{PhysRevA.76.052314}, and optical frequency standards~\cite{Ludlow2015}. Information can be stored in the internal states of a single ion and also altered, using lasers to address different electronic transitions. Cooling the ion motion is an essential part of experiments involving trapped ions, where it reduces motion-induced decoherence. It is also crucial for experiments on the interaction of light and trapped atomic systems in which the spatial spread of the atom or ion has to be reduced well below the spatial spread of a focused light beam~\cite{piro2011,hetet2013,leong2016,alber2017}. Doppler cooling is typically used to cool the ion to temperatures close to the Doppler limit~\cite{wineland1979, leibfried2003quantum, budker2004atomic}, with more sophisticated methods such as resolved sideband cooling~\cite{PhysRevLett.62.403} or EIT cooling~\cite{Morigi2000, Roos2000} being employed to cool the ion down to its ground state of motion.

Another cooling method is feedback cooling. It is well established and nowadays used routinely to cool massive oscillators (see e.g. refs.~\cite{Aspelmeyer2014, Gieseler:21} for reviews). To date, this technique has rarely been used to cool ions. In refs.~\cite{PhysRevLett.96.043003, PhysRevLett.110.133602} information about the motion of the ion has been retrieved by incorporating it into an interferometer, in which photon emission in opposite directions is superposed. A feedback signal has been generated from the recorded interference signal and applied to the electrodes of the radio frequency trap hosting the ion. As a result, the ion temperature has been reduced by 30\,\% compared to using only Doppler cooling ~\cite{PhysRevLett.96.043003}, resulting in a temperature that is 1.23 times above the Doppler limit of \(\hbar\Gamma/2 k_\mathrm{B}\). This feedback method has also recently been applied to cool the motion of a charged nanoparticle in a Paul trap~\cite{dania2022mirror}. In the field of optomechanics, various feedback schemes rely on imaging the scattered light onto a split detector, inferring position changes from the imbalance of the photo-signals. This has been successfully demonstrated for optically trapped charged nanoparticles~\cite{PhysRevLett.122.223601, PhysRevLett.122.223602} and also for a charged particle in a Paul trap~\cite{Dania2021}.

In this paper, we report on the first time that this method is applied to a single trapped ion. Compared to the method used in experiments on trapped nanoparticles in refs.~\cite{PhysRevLett.122.223601, PhysRevLett.122.223602, Dania2021}, we introduce some modification. There, the detected signal is a superposition of the light illuminating the trapped particle and the – much weaker – light scattered elastically by the particle. In experiments with single trapped ions, the cooling beam is typically focused with small NA lenses. This would lead to a poor overlap of the scattered photons with the excitation light, reducing the detectable signal. Additionally, the amount of elastically scattered light is not dominating the atomic response, except for very low saturation of the cooling transition. This is fundamentally different from the scenario found in optomechanics, where the scattering of light is entirely elastic, further reducing the applicability of the optomechanics approach. Moreover, the propagation of the cooling laser into the detection optics typically has to be avoided in experiments with trapped ions, prohibiting this method of cooling. Therefore, in our experiments we image only the fluorescence photons emitted by the ion onto a knife edge, splitting the light onto two single photon detectors. Furthermore, the feedback signal is not generated from balanced detector signals but from the signal of a single detector only. The variation of the detector signal after the knife edge is proportional to the variation in the ion’s position and can thus be used to generate a feedback signal. This signal is then processed and applied to electrodes in the vicinity of the ion, counteracting excursions from the minimum of the trapping potential. In doing so, we combine the single-detector scheme of ref.~\cite{PhysRevLett.96.043003} with the simplicity of spatially resolved detection, resulting in a feedback cooling method for ions that we believe to be applicable with low resource overhead in other experiments.

Beyond its conceptual simplicity, the method demonstrated here offers several advantages. As the experiments will show, the method can also be applied at high saturation of the ion with practically no detrimental influence on the achievable temperature. It therefore could be applied during e.g. the detection  of the internal spin state of the ion in quantum information settings, leading to a reduced upper bound for the temperature during readout and thus a reduction in cooling time after state detection. Although not investigated here, the scheme could also be used to detect excess micromotion by monitoring spectral features at the drive frequency of the ion trap. This could lead to a further method of detecting excess micromotion~\cite{keller2015}. Especially in combination with photon-correlation measurements~\cite{berkeland1998} it could provide micromotion detection capabilities in directions not covered by the cooling laser. Furthermore, we also show that by measuring the signal reflected at the ``knife edge'', and applying an additional calibration step~\cite{PhysRevLett.110.133602}, one can in parallel determine the temperature associated with the motion~\cite{Gieseler:21}. In the absence of feedback, we measure temperatures in accordance with Doppler cooling theory.

\section{Experimental setup}
\begin{figure}
    \centering
    \includegraphics[width=\linewidth]{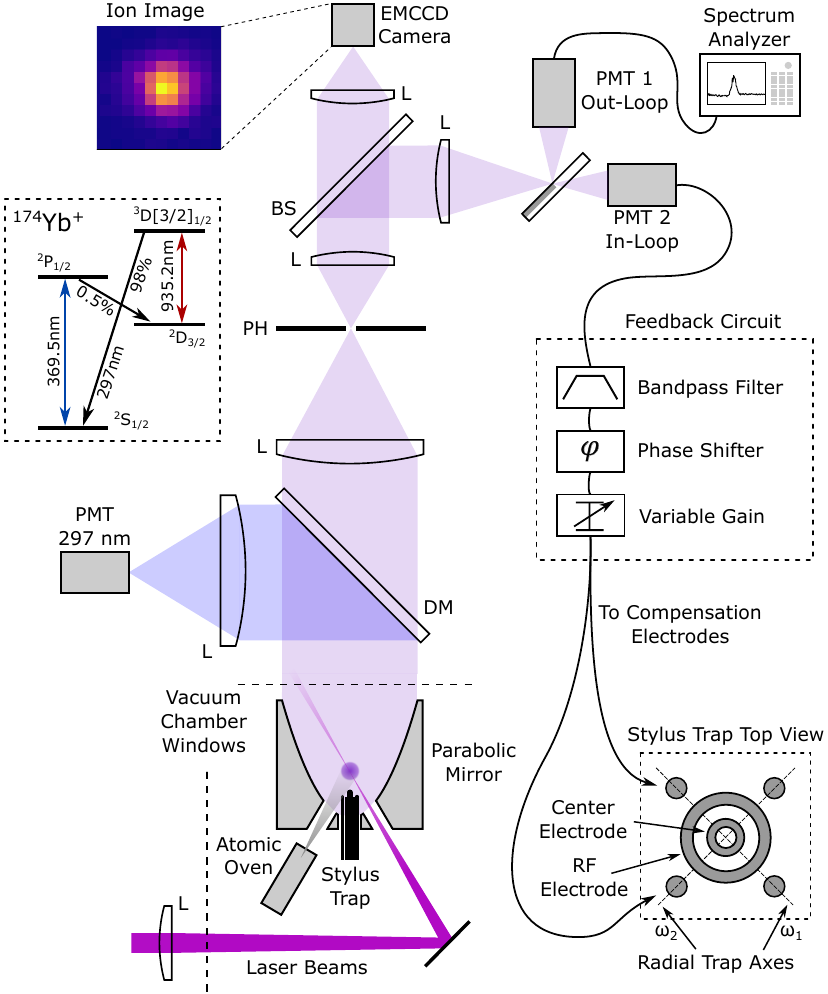}
    \caption{Schematic of the experimental setup. Fluorescence photons emitted by the single \(\mathrm{^{174}Yb^+}\) ion at 369.5\,nm are collected by the parabolic mirror, passed through a pinhole to suppress stray light, and subsequently focused onto a partially coated glass plate, which acts as a knife edge, before reaching two photomultiplier tubes (PMT). The ion motion creates an intensity modulation in the PMT signals. PMT 1 is used to measure the out-loop spectrum for thermometry, while PMT 2 is connected to the feedback circuit(s) to generate feedback signals. Each signal is applied to a different compensation electrode to independently feedback cool ion motion along either of the two radial trap axes. L: lens, BS: 95:5 (R:T) beam splitter, DM: dichroic mirror, PH: pinhole.}
    \label{fig:trapping_detection_setup_extended}
\end{figure}

For our experiments (see Fig.~\ref{fig:trapping_detection_setup_extended} for a schematic of the setup), we use a single \(\mathrm{^{174}Yb^+}\) ion which is trapped in the focus of a parabolic mirror using a stylus-like trap~\cite{maiwald2012collecting}. The procedure for trapping and Doppler cooling ions is described in~\cite{maiwald2012collecting}. The trap is mounted on a piezo-stage and can thus be moved. In order to position the ion to the focus of the parabolic mirror, the imaging system (including a filter pinhole) was first aligned with a radially polarized Laguerre-Gaussian beam passing through the focus of the parabolic mirror. After trapping, the trap assembly is moved until the image of the ion is focused onto the same location on the camera as for the alignment beam. To determine the saturation parameter of the ion for the measurements, a saturation curve of the ion is recorded as described in more detail later.

\subsection{Monitoring ion motion}
The wide open optical access of the stylus-like trap~\cite{maiwald2009} and the nearly full solid-angle coverage provided by the deep parabolic mirror~\cite{maiwald2012collecting} allow us to detect approximately 7\% of the fluorescence photons emitted by the ion. To implement motion detection, the ion is imaged onto a partially coated glass plate that acts as a knife edge. One half of the glass plate has been coated with a reflective layer of aluminum, while the other half has been left uncoated to create a sharp edge. This knife edge is aligned to on average equally distribute the fluorescence light onto two photomultiplier tubes (PMTs) when the ion is positioned in the focus of the parabolic mirror. This allows us to also use the light that is reflected from the coated half of the glass plate for monitoring the ion motion and thermometry, while the transmitted portion is used to generate the feedback signal. Using one of the PMTs solely to measure temperature effectively reduces the collection efficiency for feedback generation by half. If one, however, isn't interested in temperature measurements it is also possible to use both detectors for feedback cooling by phase shifting the signal of one detector by \(\pi\) and adding the two signals.

The harmonic oscillation of the ion in the trap along the radial trap axes causes the point-like origin of the fluorescence light to spatially oscillate across the knife edge at the corresponding trap frequencies. The intensity of the light, which either passes the knife edge or is being reflected, is thus modulated at the motional frequency of the ion. The ion motion can be directly observed in the spectrum of the signal from either detector with a spectrum analyzer. For our setup, the radial trap frequencies, perpendicular to the optical axis of the parabolic mirror, are approximately \(\omega_1 = 2\pi \cdot 450\)\,kHz and \(\omega_2 = 2\pi \cdot 455\)\,kHz, while the axial trap frequency is approximately \(\omega_3 = 2\pi \cdot 900\)\,kHz, at an RF drive frequency of \(2\pi \cdot 5.02\,\mathrm{MHz}\). In Fig.~\ref{fig:all_meas_fb_calib}(c), the ion motion along the radial trap axis with frequency \(\omega_2 = 2\pi \cdot 455\)\,kHz is clearly visible in the spectrum as a Lorentzian profile above the shot-noise background.

The angle of the knife edge relative to the radial trap axes determines the signal strength measured in the spectrum. The orientation of the radial trap axes in our setup is found by imaging the ion on an EMCCD camera and forcing it to oscillate along the trap axes. This process is described in more detail in appendix~A. By orienting the knife edge to be perpendicular to a given radial trap axis, the signal-to-noise ratio for the detected ion motion along that axis can be maximized for best feedback cooling performance. In this orientation, however, it is not possible to detect motion along the other radial trap axis, since no modulation of the light intensity occurs when the focal spot of the collected fluorescence light spatially oscillates along the knife edge. In order to observe and provide feedback cooling to both radial trap axes, the knife edge can be simply oriented such that both trap axes have a non-zero projection onto the knife edge.

Since in our experimental setup the axial trap axis is aligned parallel to the optical axis of the parabolic mirror, the knife edge only allows us to detect ion motion perpendicular to this axis, i.e. in radial direction. In experimental setups with a lens-based imaging method, it is more readily possible to detect all motional modes of the ion simultaneously. By spectrally filtering the signal, one can provide feedback cooling to all modes of motion simultaneously using only the knife edge, given suitably located electrodes. For this purpose, the lens needs to be located such that its optical axis has a non-zero projection on all motional modes of interest with the knife edge being oriented accordingly. In principle, this can also be extended to ion crystals with more complex motional modes.

\subsection{Generating the feedback signal}
In our experiment, we use the signal from the ``in-loop'' PMT, which is located in the transmission path of the knife edge, to generate the feedback signal for feedback cooling. The ``out-loop'' PMT, which is in the reflected path of the knife edge, is used as an independent detector to measure the temperature of the ion.

In order to generate the feedback signal for feedback cooling, the in-loop PMT signal is passed through a narrow Butterworth bandpass filter, with a center frequency of 455\,kHz and a 3\,dB bandwidth of 80\,kHz, to extract the ion motion. The filtered signal is then phase-shifted and amplified with a low-noise amplifier before being sent to one of the compensation electrodes located near the ion to provide the damping force. Two independent feedback circuits whose outputs are connected to two separate compensation electrodes of the trap assembly are necessary to feedback cool both radial trap axes simultaneously, since the ion motion along each axis requires in principle different phase-shift and gain settings to achieve optimal feedback cooling. In order to reduce electric noise, low-pass filters with a cut-off frequency of 70\,kHz are located close to the trap electrodes. Passing the feedback signal through these filters leads to attenuation which can be compensated for by increasing the amplification of the feedback circuit. It additionally makes the setup impervious to other noise sources (outside of the spectral range of the filters) such as laser intensity noise. As evident from the measurements (see Fig.\ref{fig:fb_cooling_temperatures}) shown later, the feedback rates are not limited by the bandwidths of these filters.

\section{Temperature measurement and feedback}
For motion along a single dimension, the power spectral density of the ion motion \(S(\omega)\) depends on the ion temperature \(T\) and is given by~\cite{PhysRevLett.96.043003, PhysRevLett.110.133602} 
\begin{equation}
    S(\omega) = \frac{4 k_B T}{m} \frac{\gamma_j}{\left(\omega^2 -\omega_j^2\right)^2 + \gamma_j^2 \omega^2}\ ,
\label{eq:psd_ion_motion}
\end{equation}
where \(m\) is the mass of the ion, \(\omega_j\) is the eigenfrequency along the trap axis \(j \in \{1,2\}\), and \(\gamma_j\) is the respective damping rate of the ion motion.

\subsection{Thermometry and feedback cooling}
By calibrating the spectrum of the detected photocurrent (see appendix~B), it is possible to perform thermometry based on the measured spectrum by fitting Eq.~\ref{eq:psd_ion_motion} to the calibrated power spectral density and accounting for an offset in the power spectral density related to the ground state motion. In addition to the measured temperature, this also yields the \(1/e\) cooling time, given by the inverse width \(1/\gamma_{\mathrm{j}}\) of the harmonic oscillator power spectral density~\cite{PhysRevLett.122.223601,PhysRevLett.96.043003,leibfried2003quantum,Dania2021}. Unless stated otherwise, the thermometry measurements were taken at a saturation parameter of $s \approx 1$. The spectrum analyzer was set to a frequency span of 50\,kHz, with a resolution bandwidth of 1\,Hz and a video bandwidth of 10\,Hz, leading to an acquisition time of approximately 4\,s. Given that the width of the motional peaks at optimal feedback cooling gain is on the order of 1\,kHz, it is easily possible to decrease the time required for temperature measurements to well below 1\,s.

The projection of a given trap axis onto the knife edge determines the signal-to-noise ratio (SNR) with which motion along that axis can be detected. Here, the cooling performance was examined for two different orientations of the knife edge, A and B, relative to the trap axes. Orientation A aims to provide the best cooling performance for ion motion along one particular trap axis. For this purpose, the knife edge is oriented to be perpendicular to the given trap axis to maximize the SNR. Orientation B sacrifices some cooling performance but allows ion motion along multiple trap axes to be feedback cooled simultaneously. Here, the knife edge is oriented such that both radial trap axes have an approximately equal projection onto the knife edge. This results in a similar SNR for ion motion detected along both axes, which is, however, lower compared to orientation A. To effectively cool motion along each trap axis and minimize crosstalk, the corresponding feedback signal is applied to a compensation electrode located along the direction of the axis.

\begin{figure}
    \centering
    \includegraphics[width=\linewidth]{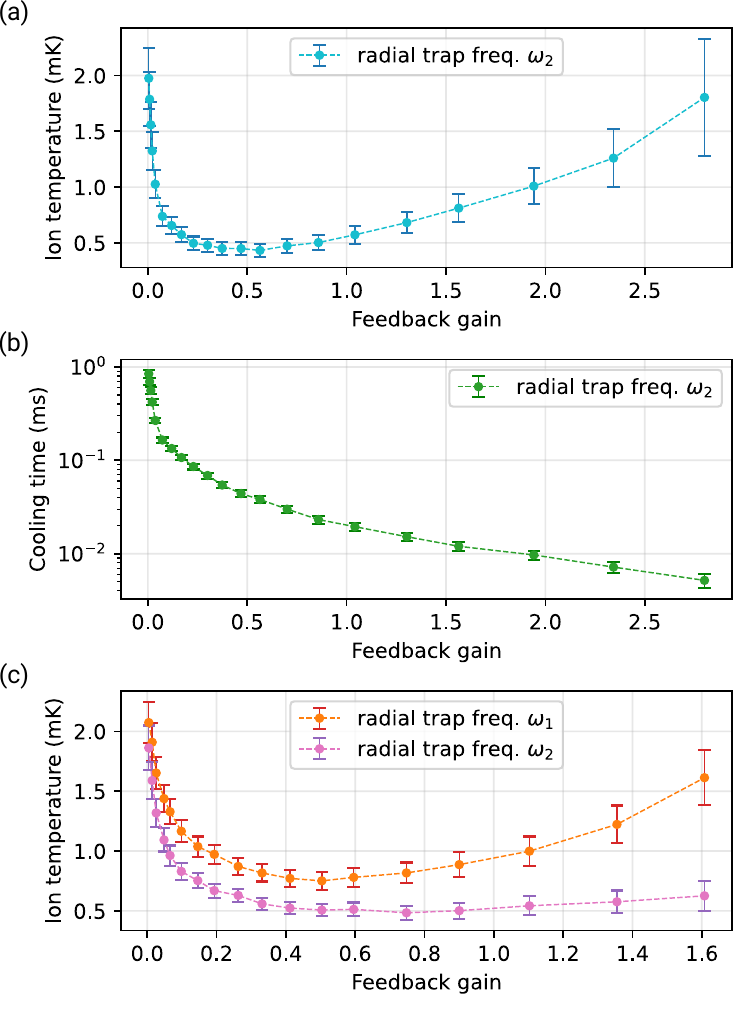}
    \caption{Temperature of the ion motion measured at \(s \approx 1\) for different feedback gain settings and knife edge orientations. (a) Orientation A provides best feedback cooling performance for ion motion along one particular trap axis. The lowest temperature achieved was \(T_{\mathrm{min},\,\omega_2} = 432 \pm 56\,\mu\)K. (b) The cooling times extracted from the linewidths of the motional spectra for orientation A show an improvement of cooling time of more than 20 at optimal feedback gain and more than 160 at the highest gain, while the temperature still remains below the temperature for pure Doppler cooling. (c) Orientation B allows ion motion along multiple trap axes to be feedback cooled simultaneously at the cost of some cooling performance. The lowest temperature achieved was \(T_{\mathrm{min},\,\omega_1} = 751 \pm 73\,\mu\)K and \(T_{\mathrm{min},\,\omega_2} = 484 \pm 52\,\mu\)K for the respective radial trap axes. The lines between data points are there to guide the eye.}
    \label{fig:fb_cooling_temperatures}
\end{figure}

\subsection{Experimental results}
Fig.~\ref{fig:fb_cooling_temperatures} shows, for both orientations of the knife edge, the temperature of the ion motion measured along the respective radial trap axes for different feedback gain settings of the feedback circuits. For orientation B, both trap axes were feedback cooled simultaneously using the same gain setting in both feedback circuits. The feedback phase was individually adjusted to provide optimal cooling. When comparing the temperatures measured for both orientations A and B, it is evident that they show an overall similar behavior.

With zero feedback gain, i.e. only Doppler cooling, the ion temperature is determined to be \(1.98 \pm 0.27\)\,mK for orientation A, with similar temperatures measured for orientation B as well. This is in good agreement with the expected temperature for our setup, which for a saturation parameter of \(s = 1\) is \(T_{\mathrm{D},\,s=1} = 1.95\)\,mK\,\cite{srivathsan2019measuring, leibfried2003quantum}.

When applying feedback, the temperature of the ion decreases with increasing feedback gain until it reaches a minimum temperature. Increasing the feedback gain further causes the ion temperature to increase again, as the applied feedback signal becomes increasingly dominated by noise~\cite{PhysRevLett.96.043003}. For orientation A [Fig.~\ref{fig:fb_cooling_temperatures}(a)], a minimum temperature of \(T_{\mathrm{min},\,\omega_2} = 432 \pm 56\,\mu\)K is reached at a feedback gain of 0.56, which is \(77.8 \pm 2.9\)\,\% below the temperature without feedback. Furthermore, the minimum temperature achieved in this feedback configuration reaches the Doppler limit temperature~\cite{wineland1979, mandel1995optical, budker2004atomic} of \(T = \hbar \Gamma / 2 k_\mathrm{B} = 470\,\mu\)K for the linewidth \(\Gamma = 2\pi \cdot 19.6\,\mathrm{MHz}\) of the transition used for Doppler cooling. However, one has to keep in mind that the Doppler limit temperature can only be reached when the cooling laser is perfectly aligned with the trap axis. In experiments where a single laser is used for Doppler cooling, it's wave vector needs to have a non-zero overlap with all three trap axes, increasing the minimally achievable Doppler temperature. For orientation B [Fig.~\ref{fig:fb_cooling_temperatures}(c)], the minimum temperature achieved for both trap axes is \(T_{\mathrm{min},\,\omega_1} = 751 \pm 73\,\mu\)K and \(T_{\mathrm{min},\,\omega_2} = 484 \pm 52\,\mu\)K at a feedback gain of 0.50 and 0.75, respectively. We attribute the difference between these two temperatures to crosstalk between the two feedback cooling signals. The two directions are affected differently due to an asymmetry arising from different orientations of the feedback electrodes relative to the trap axes. This, as well as the spectral overlap of the nearly degenerate secular modes of the radial trap frequencies, is a particular issue in our setup and might not pose an issue in other setups.

Around the minimum, only a weak dependence of the temperature on the feedback gain is observed, which is of advantage for the long-term stability of the cooling mechanism. During adjustment, it was also found that a change in the phase setting of approximately \(\pm 10^\circ\) around the optimal value only has a weak influence on the measured temperature.

It is evident from Fig.~\ref{fig:fb_cooling_temperatures}(b) that the \(1/e\) cooling time of the motion decreases with increasing feedback gain. Compared to solely using Doppler cooling with a measured cooling time of \mbox{\(\sim 1\,\mathrm{ms}\)}, additional feedback cooling at the appropriate gain setting can be used to reach the aforementioned minimum temperature with an improved cooling time of \(38.06 \pm 3.32\;\mu\mathrm{s}\). At even higher gain settings, feedback allows us to reach temperatures similar to the case of only Doppler cooling but with a \(1/e\) cooling time of \(5.18 \pm 0.90\;\mu\mathrm{s}\), resulting in a speedup of more than 160 times compared to without feedback.

\subsection{Dependence of cooling performance on saturation}
\begin{figure}
    \centering
    \includegraphics[width=\linewidth]{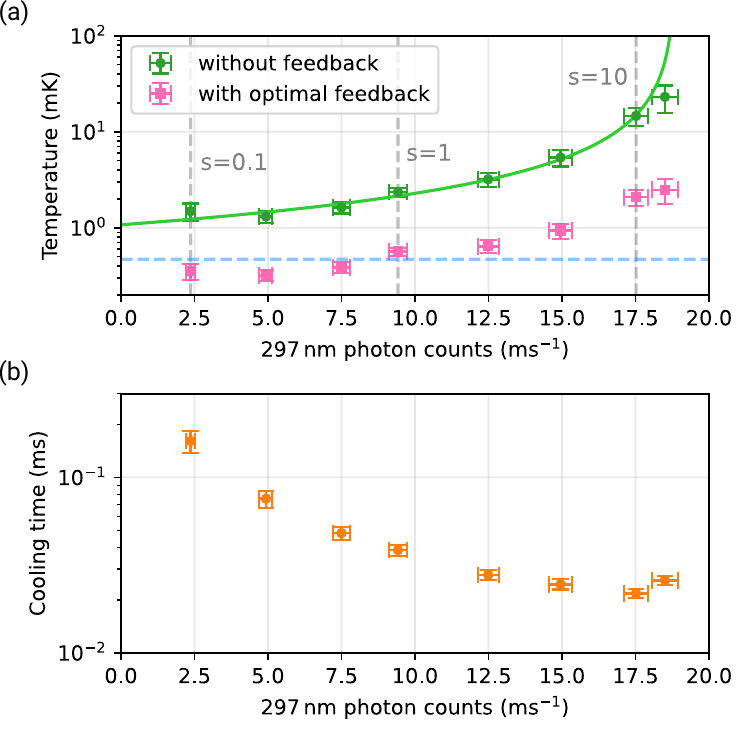}
    \caption{(a) Temperature of the ion motion along the radial trap axis with frequency \(\omega_2\) vs. different count rates of detected photons with a wavelength of 297\,nm scattered by the ion. At each count rate, a measurement similar to the one shown in Fig.~\ref{fig:fb_cooling_temperatures}(a) was taken. The solid green line shows a fit of Eq.~\ref{eq:ion_temp_count_rate} to the data without feedback cooling. The gray dashed lines indicate the count rates corresponding to different saturation parameters, parametrized by saturation measurements. The blue dashed line indicates the temperature \(T = \hbar \Gamma / 2 k_\mathrm{B} = 470\,\mu\)K. Panel (b) shows the \(1/e\) cooling time at the feedback gain where the temperature shown in panel (a) was reached.}
    \label{fig:ion_temp_sat_para}
\end{figure}

In order to investigate the achievable feedback cooling performance at different saturation parameters, the measurement previously shown in Fig.~\ref{fig:fb_cooling_temperatures}(a) for orientation A was repeated for different cooling laser powers, and thus photon scattering rates. The temperature without feedback, as well as the minimum temperature achieved with feedback cooling, were noted for each measurement and are shown in Fig.~\ref{fig:ion_temp_sat_para}.

To determine the corresponding saturation parameter \(s\) of the ion in each measurement, the rate of detected 297\,nm photons is used, which are scattered by the ion on the \(\mathrm{^3D[3/2]_{1/2} - ^2S_{1/2}}\) transition.  The number of photons scattered on this transition is proportional to the number of photons scattered on the \(\mathrm{^2S_{1/2} - ^2P_{1/2}}\) transition at a wavelength of 370\,nm used for Doppler cooling. However, detecting photons with a wavelength of 297\,nm allows us to spectrally filter out all laser stray light from the cooling lasers used in the setup. In order to make the data points more easily comparable, we keep the detuning fixed at half linewidth detuning. The rate of scattered photons depends on the saturation parameter and follows the relation \(R_\mathrm{297} \sim s/(1 + s)\). Given that the temperature of the ion also depends on the saturation parameter via \(T_\mathrm{D} \sim (1 + s)\)~\cite{leibfried2003quantum} (when the detuning of the cooling laser is kept fixed at the optimal detuning of \(-\Gamma/2\), for low saturation parameters, and not increased by a factor of \(\sqrt{1+s}\)), the ion temperature can be expressed in terms of the detected 297\,nm photon count rate as
\begin{equation}
    T_\mathrm{D}(R_{\mathrm{297}}) = T_{\mathrm{D},\,s=0} \left(1 + \frac{R_{297}}{R_{297,\,\mathrm{max}} - R_{297}} \right),
    \label{eq:ion_temp_count_rate}
\end{equation}
where \(T_{\mathrm{D},\,s=0}\) is the temperature achieved with Doppler cooling at \(s = 0\) for the geometry of our setup and \(R_{297,\,\mathrm{max}}\) is the maximum possible rate of 297\,nm photons scattered by the ion. Fitting this equation to the temperatures without feedback (shown in green in Fig.~\ref{fig:ion_temp_sat_para}) yields \(T_{\mathrm{D},\,s=0} = 0.99 \pm 0.05\,\mathrm{mK}\), and a maximum 297\,nm count rate of \(R_{297,\,\mathrm{max}} = 18.88 \pm 0.33\,\mathrm{ms^{-1}}\). This agrees well with the maximum count rate of \(18.58 \pm 0.10\,\mathrm{ms^{-1}}\) for \(s \rightarrow \infty\), deduced from separate saturation measurements. The Doppler temperature limit \(T_{\mathrm{D},\,s=0}\) determined in this measurement agrees with the previously calculated value \(T_{\mathrm{D},\,s=1} = 1.95\)\,mK, given that \(T_{\mathrm{D},\,s=1} = 2 \cdot T_{\mathrm{D},\,s=0}\).

Comparing the temperatures without feedback cooling with the minimum temperatures achieved with additional feedback, it is evident that feedback cooling allows the ion to remain at significantly reduced temperatures. For low saturation parameters (and hence, low number of scattered photons at a wavelength of 297\,nm) it is even possible to cool the ion to temperatures below the theoretical Doppler limit of \(\hbar\Gamma/2\mathrm{k}_\mathrm{B}\). At the lowest point, we measure a temperature of \(319.8 \pm 43.6\,\mu\mathrm{K}\). The relative temperature reduction with feedback cooling becomes especially pronounced for higher saturation parameters. This, together with \(1/e\) cooling times of a few tens of \(\mu\mathrm{s}\), might be beneficial for quantum computing applications, as this enables the ion's internal spin state to be read out at high photon scattering rates while still remaining comparatively cool. Additional noise, that might arise from the feedback circuit, can simply be avoided during gate operations by switching the feedback circuit off during these operations.

\section{Concluding discussion}
We have demonstrated that it is possible to measure the spectrum of a trapped ion's motion by focusing the collected fluorescence light onto a knife edge and detecting the transmitted or reflected photons.  This is in contrast to other methods based on imaging the ion onto a camera and integrating over time scales much larger than the oscillation periods of the ion in the trapping potential~\cite{blinov2006, norton2011, knuenz2012, rajagopal2016, srivathsan2019measuring}. Calibrating these spectra allows us to perform absolute temperature measurements.

By deriving a feedback signal from the measured ion motion and applying it to a set of nearby electrodes, we achieved a reduction in ion temperature down to and even slightly below the theoretical Doppler limit of \(\hbar \Gamma / 2 k_\mathrm{B}\). We could show that the method presented here leads to a reduction of the \(1/e\) cooling time to the order of \(10-100\,\mu\mathrm{s}\). We have furthermore demonstrated that changing the orientation of the knife edge enables the simultaneous detection and cooling of the motion along all directions orthogonal to the optical axis along the imaging path.

One might wonder which temperatures could in principle be achieved with feedback cooling. For feedback cooling in optomechanics experiments, the achievable temperature or corresponding mean number of phonons, respectively, is given by $n=(1/\sqrt{\eta}-1)/2$~\cite{rossi2018}, where $\eta$ is the efficiency of measuring the motion of the ion. 
Approximating $\eta$ with the detection efficiency of the out-of-loop detector of 0.035 would yield $n\approx2$, which is well below the number of 15 phonons corresponding to the lowest temperature of $320\,\mu\text{K}$ observed in our experiments. 
We attribute this to the influence of the simultaneous Doppler cooling, where recoil heating due to photon scattering is the dominant heating mechanism. The influence of recoil heating increases at increasing saturation parameter, which is compatible with the observation made in Fig.~\ref{fig:ion_temp_sat_para}(a) that the minimum temperature achievable with feedback cooling increases with larger saturation parameters.

The relative simplicity of the setup leads to a high long-term stability, resulting in constant feedback gain and phase settings over several hours. This, together with the fast cooling times and relatively low achievable temperatures, might proof useful for quantum computing in view of the recent progress of high fidelity two-ion gates that don't require ground state cooling~\cite{hughes2025} or any other scenario in which intermediate temperatures below the ones obtained with Doppler cooling are sufficient.

\section*{Acknowledgment}
This research was conducted within the scope of the project QuNET, funded by the German Federal Ministry of Science, Technology, and Space (BMFTR) in the context of the federal government’s research framework in IT-security ``Digital. Secure. Sovereign'' (Grant 16KIS1264).


\appendix

\section{ORIENTATION OF THE RADIAL TRAP AXES}
\begin{figure*}
    \centering
    \includegraphics[width=\linewidth]{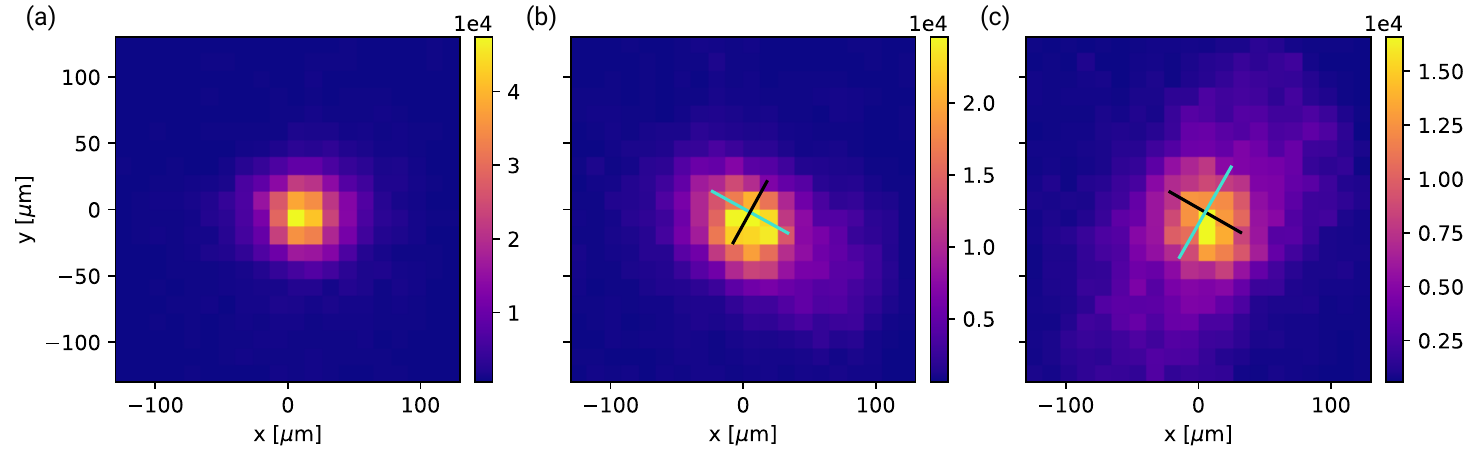}
    \caption{Images of the ion recorded with an EMCCD camera for different setups of the external drive to determine the orientation of the radial trap axes. (a) Ion without external drive applied. (b) and (c) Ion with external drive applied to resonantly drive the ion along the radial trap axis with frequency \(\omega_1\) and \(\omega_2\) respectively. The turquoise lines indicate the radial trap axis along which the ion is driven by the external drive, while the black lines indicate the other remaining radial trap axis. The length of the lines corresponds to the width of the Gaussian intensity profile of the ion image along the given axis. The intensities in the image are given in units of the camera's analog-to-digital converter.}
    \label{fig:ion_images}
\end{figure*}

The orientation of the radial trap axes in our setup is found by imaging the ion on an EMCCD camera. An external drive is applied to different compensation electrodes of the trap assembly to force the ion to oscillate along the corresponding trap axes at the respective trap frequencies. Fig.~\ref{fig:ion_images}(a) shows an image of the ion without an applied external drive as a reference. When an external drive is applied, the ion appears elongated along the respective trap axis because of the induced oscillatory motion of the ion and the comparably long exposure time of the camera with respect to the oscillation period [Fig.~\ref{fig:ion_images}(b) and Fig.~\ref{fig:ion_images}(c)]. A 2-D Gaussian intensity profile is fitted to the images to determine the orientation of the radial trap axes. The angle between the radial trap axes with frequency \(\omega_1\) and \(\omega_2\) and the horizontal axis (x-axis) in the camera images was found to be \(\alpha_1 = -28.87^\circ \pm 0.09^\circ\) and \(\alpha_2 = 60.24^\circ \pm 0.06^\circ\), respectively. The angle between the trap axes and the knife edge can then be calculated accordingly for the calibration measurements described in appendix~B.

\section{CALIBRATION OF THE SPECTRUM}
In order to measure \(S(\omega)\), we first consider the photocurrent that is induced on a detector by the fluorescence light that passes the knife edge. The parabolic mirror and the imaging system image the ion onto the knife edge with a given magnification \(M\). For small displacements of the ion inside the parabolic mirror, the intensity distribution of the light focused onto the knife edge can be approximated with a Gaussian intensity distribution of size \(\sigma\). The knife edge is positioned such that on average half of the impinging fluorescence light passes to maximize the sensitivity of the setup to ion motion. An error function can thus be used to describe the intensity of light passing the knife edge. The motion of the ion \(x(t)\) in the trap potential then effectively causes the mean position of the Gaussian focal spot to vary across the knife edge. The photocurrent \(\mathrm{i}(t)\) induced on a detector placed after the knife edge by this intensity distribution is thus given by
\begin{equation}
\tag{B1}
    \mathrm{i}(t) = \frac{\mathrm{i_0}}{2} \left[1 + \mathrm{erf}\left(\frac{M\,x(t)}{\sigma \sqrt{2}}\right)\right] \approx \frac{\mathrm{i_0}}{2} \left[1 + \frac{M\,x(t)}{\sigma \sqrt{2}}\right],
\end{equation}
where \(\mathrm{i_0}\) is the average intensity of the fluorescence light incident on the knife edge. For small values of \(x(t)\), which are typically the case in the experiment, the response of the error function is approximately linear.

\subsection{First measurement}
The calibration method presented here consists of two measurements. For the first measurement, the response of the detector signal/photocurrent on a displacement of the ion is measured. In the experiment, the dependence of the signal from the out-loop PMT on the position of the ion is determined by moving the ion perpendicular to the optical axis of the parabolic mirror. This is achieved by moving the trap using the piezo translation mount and recording the detected photon count rate at each position. The resulting count rates are then normalized to the count rate measured when the ion is positioned at the focus of the parabolic mirror. The slope of the linear region \(\Delta d\) of the normalized count rate then allows us to translate a change in the measured count rate of the detector to a corresponding displacement of the ion. This slope was measured for two different orientations of the knife edge relative to the trap axes. Note that it is also possible to move the knife edge instead of the ion to determine the slope and the magnification \(M\) independently.

In the case of the knife edge being oriented perpendicular to one given radial trap axis, in order to achieve best feedback cooling performance, the measurement is performed by moving the ion along the diagonal direction [Fig.~\ref{fig:all_meas_fb_calib}(a)]. The resulting slope of the normalized count rate per unit of ion displacement \(\Delta d = (4.46 \pm 0.14)\,\mu\mathrm{m}^{-1}\), with the projection of the movement direction onto the trap axis accounted for, is then used for the calibration of that particular experimental configuration.

When the knife edge is oriented such that both radial trap axes have a non-zero projection onto the knife edge to feedback cool both trap axes simultaneously, the ion is simply moved perpendicular to the knife edge to measure the change in the detected count rate for a given displacement. The effective slope for each trap axis is then calculated from the resulting slope by taking into account the angle between the axes and the knife edge.

It should be noted, that the linear region here amounts to approximately half a \(\mu \mathrm{m}\), governed by the imaging point spread function of the setup. Compared to experiments relying on the interference of the light scattered by the ion with itself, where the linear region is on the order of \(\lambda/4\), this leads to a reduced sensitivity on displacements of the trapping region.

\subsection{Second measurement}
\begin{figure*}
    \centering
    \includegraphics[width=\linewidth]{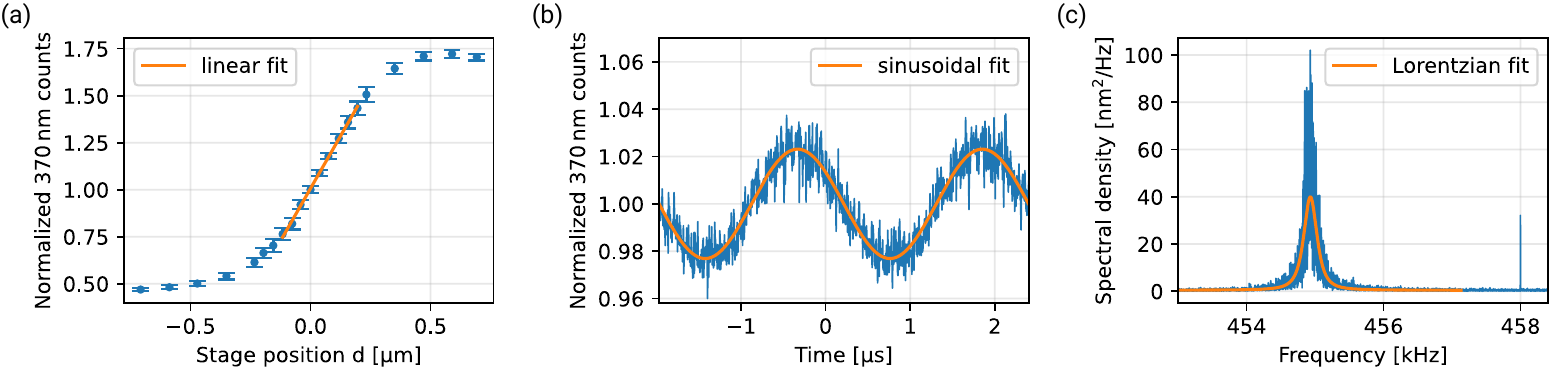}
    \caption{(a) Normalized count rate of 369.5\,nm photons measured at different positions of the ion trap along the trap axis with frequency \(\omega_2\). The count rates are normalized to the value measured when the ion is positioned in the focus of the parabolic mirror, i.e. \(d = 0\). (b) Normalized correlation between detected 369.5\,nm photon counts and an external sinusoidal drive with a frequency of 458\,kHz applied to the ion. The drive frequency was chosen such that the frequency is close to, but still several linewidths away from the secular motion. (c) Calibrated power spectral density of the signal from the out-loop PMT showing ion motion along one radial trap axis at a center frequency of \(\omega_2 = 2\pi \cdot 455\)\,kHz, as well as the external drive used to calibrate the spectrum.}
    \label{fig:all_meas_fb_calib}
\end{figure*}

For the second measurement required for the calibration, an external sinusoidal AC electric drive field is applied to one of the compensation electrodes (cf. ref.~\cite{PhysRevLett.110.133602}) to force the ion to coherently oscillate with a fixed amplitude at a frequency close to but not overlapping with the spectral feature of the secular motion.
This results in a modulation of the detected photon count rate owing to the motion of the ion at the drive frequency. The modulation can be recorded by correlating the time delays between the detection events and the zero-crossing of the external drive [Fig.~\ref{fig:all_meas_fb_calib}(b)].

The modulation amplitude \(A_\mathrm{corr}\), as seen in the 370\,nm count rate, is directly related to the ion displacement amplitude \(A_\mathrm{displ} = A_\mathrm{corr} / \Delta d\) by the slope \(\Delta d\) previously determined in the first measurement. Furthermore, the coherent oscillation of the ion is visible in the spectrum of the photocurrent as a narrow peak at the drive frequency [Fig.~\ref{fig:all_meas_fb_calib}(c)]. Without an ion present in the setup, the background noise was measured to be a factor of 20 below the noise floor shown in Fig.~\ref{fig:all_meas_fb_calib}(c). The height of the spectral peak \(\Tilde{S}_\mathrm{peak}\) relates directly to the previously inferred ion displacement amplitude \(A_\mathrm{displ}\) and thus can be used to calibrate the spectrum of the detected photocurrent \(\Tilde{S}(\omega)\), measured with a given resolution bandwidth (RBW), to obtain the power spectral density of the ion motion \(S(\omega) = \Tilde{S}(\omega) \cdot A_\mathrm{displ}^2 / (\Tilde{S}_\mathrm{peak} \cdot \mathrm{RBW})\).

\bibliography{bibliography}

\end{document}